\documentclass[pss,fleqn]{w-art}
\usepackage{times}
\usepackage{w-thm}
\usepackage[]{graphicx}
\begin{document}
\DOIsuffix{theDOIsuffix}
\Volume{XX}
\Issue{1}
\Month{01}
\Year{2003}
\pagespan{1}{4}
\Receiveddate{15 July 2006}
\Reviseddate{}
\Accepteddate{}
\Dateposted{}



\title[TAMR  in Coulomb Blockade]
{Anisotropic magnetoresistance in single electron transport }


\author[J. Fern\'andez-Rossier]{J. Fern\'andez-Rossier \footnote{Corresponding
     author: e-mail: {\sf jfrossier@ua.es}, Phone: +34\,96\,590\,3400-2019,
     }\inst{1}} \address[\inst{1}]
{Department of Applied Physics, University of Alicante, 03690 Spain}
\author[]{R. Aguado
\inst{2}}
\address[\inst{2}]{  Instituto de Ciencia de Materiales, CSIC, Cantoblanco,
Madrid 28049, Spain}
\author[]{L. Brey\inst{2}}
\begin{abstract}
We study the effect of magnetic anisotropy in a single electron transistor with
ferromagnetic electrodes and a non-magnetic island.  We identify the variation
$\delta \mu$ of the chemical potential of the electrodes as a function of the
magnetization orientation as a key quantity that permits to tune the electrical
properties of the device. Different effects occur depending on
the relative size of  $\delta \mu$ and the charging energy.
We provide preliminary quantitative estimates of $\delta \mu$ using a
very simple toy model for the electrodes.

\end{abstract}
\maketitle                   


{\em Introduction}
Spin polarized transport through carefully designed ferromagnet-non magnetic
-ferromagnet (FM$_1$-NM-FM$_2$) nanostructures can be very sensitive to the
relative orientation of their magnetic moments\cite{Maekawa-book},
$\vec{\Omega}_1$ and $\vec{\Omega}_2$ . The
latter are
controlled by external  magnetic fields  that result in the so called  tunnel
magnetoresistance\cite{TMR} (TMR), Giant magnetoresistance\cite{GMR} (GMR) and
Ballistic magnetoresistance\cite{BMR} (BMR)  when the NM layer is a tunnel
barrier, a metal and a geometrical nanoconstriction respectively.

In bulk ferromagnets, the dependence of resistance on the angle between the
magnetization $\vec{\Omega}$ and the current $\vec{j}$ gives rise to the so
called anisotropic magneto-resistance (AMR). The microscopic origin of this
phenomenon is the spin
orbit interaction, which also accounts for the stability of the magnetization
orientation (magnetic anisotropy).  Recently,  the concept of Tunneling
Anisotropic Magnetoresistance (TAMR)  has been
proposed theoretically  \cite{TAMR-theory} and independently
  verified in experiments \cite{TAMR-exp,TAMR-exp2,TAMR-exp3}
 in tunnel junctions with GaAsMn electrodes.
The related concept of Ballistic Anisotropic Magneto Resistance (BAMR) has
been proposed theoretically \cite{BAMR-theory} and
 observed in atomic sized Nickel \cite{Ralph-BAMR} and Iron nanocontacts
 \cite{Viret-BAMR}. As opposed to TMR and BMR,
where the high and low resistance states are related to variations in
$\vec{\Omega}_1\cdot\vec{\Omega}_2$,
 TAMR and BAMR effects occur for
 $\vec{\Omega}_1=\vec{\Omega}_2\equiv\vec{\Omega}$ and
 depend on the angle between the transport direction and
 $\vec{\Omega}$, which is controlled by an external field.

The microscopic origin of BAMR and TAMR can be traced back to the dependence of
the electronic structure on the angle between $\vec{\Omega}$ and the  crystal
lattice, originated by the spin orbit (SO) coupling.
In ideal 1-dimensional chains,
 BAMR occurs if the number of bands at the Fermi energy is different for the
magnetization parallel and perpendicular to current flow \cite{BAMR-theory}.
 In real
metallic   nanocontacts,  the transmission of the different channels
is not perfect and
an ab-initio approach \cite{Jacob05} extended to  include SO interaction
would be necessary to account for the experimental reports.
In the case of TAMR
the relevant quantity is the transmission \cite{TAMR-theory}, which is related to
the density of states at the Fermi energy.
The size of the  AMR effects depends on
the relative ratio of the spin orbit interaction $\Delta_{SO}$ , the Fermi
energy $\epsilon_F$ and the exchange splitting $J$.  Not surprisingly, TAMR has
been reported first  in III-V ferromagnetic semiconductors, where SO coupling is
the largest energy scale in the system.

Motivated by recent experimental results \cite{Wunderlich}, we consider  a
different kind of   AMR effect, which takes place in a single electron
transistor (SET) with ferromagnetic electrodes. Although in the  
experiments\cite{Wunderlich}  both the electrodes and the dot are made of
ferromagnetic GaMnAs, here  we consider a simpler SET  with a  
non-magnetic island (NMI). To the best of our knowledge AMR effect in this
kind of SET have not been explored so far 
\cite{Ralph-CB,Ralph-Science}.  The NMI is influenced by the magnetization orientation  of the
electrodes, $\vec{\Omega}$,  both because the tunneling rates $\Gamma$
and the  electrode  chemical potential $\mu$ depend on $\vec{\Omega}$.
Whereas the change in $\Gamma$ is related to TAMR, the 
effect described here is related also to the change in chemical potential, and
therefore closer to the so called magneto Coulomb Blockade effect \cite{MCB}.

In the rest of this paper we give a {\em preliminary} account of how the
magnetization orientation of ferromagnetic electrodes affects the properties of
a SET. The latter are described with a very simple toy model with  Rashba spin
orbit and Stoner magnetism. We discuss how, depending on the ratio between the
different energy scales of
the device, different magneto-transport effects are possible.  


{\em Theory}.
We consider single electron tranport through a  NMI in
the Coulomb Blockade (CB) regime, weakly coupled to two identical ferromagnetic
electrodes, denoted by left (L) and right (R).
The reverse situation,  a magnetic dot with strong anisotropy
coupled to non-magnetic electrodes, has been considered elsewhere
\cite{JFR-Aguado06}.
The island is capacitively coupled to a gate electrode, which
is able to change electrostatic potential, and thereby the charge, inside the
NMI.  The Hamiltonian of the system reads:
\begin{equation}
H=\sum_{k,\nu,\eta=L,R} E^{\eta}_{k\nu}
\eta^{\dagger}_{k,\nu}\eta_{k,\nu}+
 \sum_{n} \left(\epsilon_n+ e \phi_{ext}\right) d^{\dagger}_{n\sigma}
 d_{n\sigma}
 + \frac{\hat{\cal Q}^2}{2C}
   +
\sum_{\sigma,k\nu,\eta}
\langle k\nu  \eta|V |\sigma n\rangle
\eta^{\dagger}_{k\nu} d_{n\sigma} + h.c.
\end{equation}
The first term describe the ferromagnetic electrodes in
a mean field approximation  that yields single particle states labeled with
$|k\nu\rangle$, where $\nu$ is a band index that includes spin. Because of spin
orbit interaction, the spin $\sigma$ of the electrons in the electrodes is not
conserved.  The next 2 terms describe the
single particle part, including the coupling to the gate potential $\phi_{ext}$,
 and the electrostatic
Coulomb repulsion of the NMI Hamiltonian
in the orthodox model \cite{Beenakker}.
 Since we neglect spin-orbit interactions in the NMI, the
single particle levels $n$ are also eigenstates of the spin operator, which
we quantize parallel to the electrode magnetic moment
orientation,  $\vec{\Omega}$. The last term describes the tunneling
of electrons between the electrodes and the NMI. This tunneling occurs via a
{\em spin conserving} operator, $V$. The
 single particle energies of the NMI
are denoted by $\epsilon_n$, and $\hat {\cal Q }=e\sum_n
d^{\dagger}_n d_n$ is NMI charge operator. $C$ is the capacitance of the
island.


The most fundammental result of orthodox CB theory states that the
number of electrons in the island changes from $N$ to $N+1$ when
the gate electrode sets into resonance the  ground state energies
with $N$ and $N+1$ electrons. In the framework of othodox CB
theory this  occurs when \cite{Beenakker}:
\begin{equation}
 \epsilon_{N+1} +E_C\left(N+\frac{1}{2}\right)
 -e \phi_{ext}
 =\mu\left(\vec{\Omega}\right)
\label{CB}
\end{equation}
where $\mu$ is the chemical potential of the left and right electrodes,
 $e\phi_{ext}$ is the gate
potential 
 and $E_C\equiv \frac{e^2}{C}$ is the SET charging
energy.
 Since the chemical potential of the electrodes depends
on the orientation of their magnetic moment, $\vec{\Omega}$, it is
apparent that the charging curve $N(\phi_{ext},\vec{\Omega})$
of the SET depends on $\vec{\Omega}$.

As the gate is ramped, peaks in the zero bias conductance
appear at the degeneracy point between ground states with $N$ and $N+1$
electrons. The width of the peaks is related to
the tunneling rates $\Gamma^{\eta}_{n\sigma}$,
 which also depend on $\vec{\Omega}$.
Since $\Gamma^{\eta}_{n\sigma}$ can be made arbitrarily small,
we can in a first stage ignore its dependence on $\vec{\Omega}$  and discuss the effects related to a change in the chemical
potential of the electrodes,
as $\vec{\Omega}$ is rotated with a magnetic
field. For the sake of the discussion we assume a uniaxial magnet, so that the
angle $ \theta$ characterizes the angle between the  magnetization and the axis
$\hat z$:
$\vec{\Omega}=\left(sin(\theta),0,\cos(\theta)\right)$.
We consider two orientations of the magnetization of the
electrodes, yielding different chemical potentials, $\mu_1=\mu(\theta=0)$ and
$\mu_2=\mu(\theta=\pi/2)$. We define $\delta \mu\equiv \mu_1-\mu_2$.
Depending on the relative value of $\delta \mu$, $\Gamma$ and $E_C$,
we  distinguish 3 different regimes:
(i) {\bf Standard}. This occurs if $\delta \mu<<\Gamma$. In this case the
effect of the change of chemical potential is negligible in both the charging
state and the conductance of the device.

(ii) {\bf Magnetoresistive}   $\Gamma<\delta \mu<<E_C$. In this case
the change of the chemical potential is big enough as to detune the single
electron device from resonance. If the gate potential is tuned
to set the device into a peak of conductance at a given $\vec{\Omega}_1$,
 a large change in the conductance will take place when the orientation of
 the electrode magnetization is changed to $\vec{\Omega}_2$ so that
 $\delta \mu>\Gamma$.   The
change in resistance can be positive or negative depending on whether the
single electron transistor gets closer or further away from the CB peak.
 In this regime the charge state of the dot is {\em not affected}, or at least
weakly affected but the dependence of $\Gamma$ on $\vec{\Omega}$ will
play an important role, as it happens in the case of TAMR.

(iii) {\bf Magnetocapacitive}   $\Gamma<<E_C<\delta \mu$ In this
case the change of the chemical potential is big enough as to {\em change the
charge state of the dot}. Neglecting the single-particle spacing, the  charging
energy for  $\delta N$ extra electrons is $\delta N E_C$. Therefore, the number
of carriers in the NMI can change by as many as
$\delta N=\frac{\delta \mu}{E_C}$.
This anisotropic magneto-capacitive effect is different from TAMR
and specific of SET with ferromagnetic electrodes.
 The number of electrons $\delta N$
injected or extracted from the NMI
could be monitored by the number of peaks in the zero bias conductance
as an external magnetic field rotates $\vec{\Omega}$ so
that $\mu$ crosses the SET charging boundaries defined by (\ref{CB}).

Whereas there is plenty of experimental information about typical charging
energies  of SET, which  range from $\mu eV$ to meV, we are not aware of reports
of the dependence of  chemical potential on magnetization
orientation. Calculations of this quantity using realistic models or ab-initio
 are necessary. Here we use a very simple model to explore this problem.
 We describe the  electrodes as a one dimensional electron gas
with parabolic dispersion and spin splitting $J$   modified by
  a Rashba spin
orbit term:
$H_{el } = \frac{p^2}{2m} \delta_{\sigma,\sigma'} -
\frac{J}{2} \vec{\sigma}\cdot\vec{\Omega}
+ \lambda p\sigma_z$.
Here $m$ is the effective mass (in units of the free electron mass)
and $\lambda$ is
the strength of the spin-orbit interaction. We take $\hbar=1$.
and  $\lambda$ has dimensions of velocity.  A dimensionless ratio of Stoner and spin orbit interactions is
given by $(\lambda \hbar k_f)/J$, where $k_f$ is the Fermi wave-vector.
In magnetic semiconductors like GaMnAs the
exchange splitting is comparable or even smaller than the SO coupling. In metals
like Nickel the exchange splitting is more than 10 times larger than the SO
orbit interaction.

The eigenstates $|k\nu\rangle$ and eigenvalues $E_{k\nu}$ of
$H_{el }$ can be obtained  analytically. In figure 1 we plot
$E_{k\nu}$ for $J=1$, $m=\lambda=0.5$. For this rather large value
of the spin orbit interaction, the bands change significantly as
the $\vec{\Omega}$ rotates. In this sense the situation is similar
to the case of GaMnAs \cite{Abolfath}
 Upon
simple quadratures we can obtain $\mu(\vec{\Omega})$
 for fixed values of $J$,$m$ and $\lambda$ as a function of the electron
 density.
In fig. 1d  we see that the $\mu(n)$ curve is different for different values of $\theta$.
  The jump in $\mu(n)$ is related to the occupation of the upper band,
  which occurs at a  value of the density  that  depends on $\vec{\Omega}$.
   Assuming a constant
 electronic density we can invert the curve $\mu(n)$ and obtain $\delta \mu$
 (fig. 1e).
  We see that $\delta \mu$ can be either positive or negative, depending on $n$, and its
 absolute value ranges 0 and  0.15 $J$, which would largerly exceed the change
 in $\nu$ when the magnetic field is applied parallel to $\vec{\Omega}$
 \cite{MCB}.  The change in size and magnitude of
 $\delta \mu$ as a function of $n$ are related to the change of the ratios between exchange energy,
 band energy and spin orbit coupling.  In fig. 1f and 1g we show $ \mu(\theta)$
 for $n=0.5$ and $n=0.3$ respectively. The horizontal lines in 1g stand
 for values of $\mu$ at which the number of electrons in the NMI changes,
 assuming $E_C=0.05J$,  we see how the dot would gain up to 2 extra electrons as the $\theta$ is varied
 from $\pi/2$ (the easy axis in this case) to 0 or $\pi$. The vertical lines in
 1g stand for the values of $ \theta$ at which a conductance peak would be
 observed.

 \begin{figure}
[t]
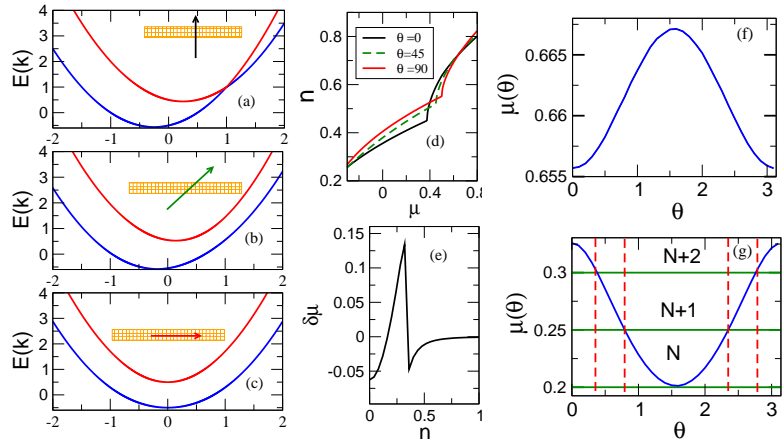

\includegraphics[width=2.5in]{FIGURE2b.eps}
\includegraphics[width=1.5in]{FIGURE3b.eps}
\caption{ \label{fig1}. {\bf 1a-1c}: Bands for three different
magnetization orientations, $\theta=0$ (a), $\theta=\frac{\pi}{4}$
(b) and $\theta=\frac{\pi}{2}$ (c). {\bf 1d}: density vs chemical
potential for the 3 same values of $\theta$. {\bf 1e}: difference
in chemical potential for $\mu(0)-\mu(\pi/2)$.{\bf 1f,1g}
$ \mu(\theta)$ for $n=0.5$ (f) and $n=0.3$ (g). In the latter we
also show the charge in the dot for different values of $\mu$.  }
\end{figure}



We now discuss briefly how the tunneling rates are affected by the
rotation of the electrode magnetization. The scattering rate for the $n$
level of the QD, with spin $\sigma$ along the $\vec{\Omega}$ axis reads:
\begin{equation}
\Gamma^{\eta}_{n\sigma}(\vec{\Omega})= \frac{1}{\hbar}
  \sum_{k,\nu=\pm} \left|\langle k \eta |V|n\rangle\right|^2
|{\cal C}_{k,\nu,\sigma}(\vec{\Omega})|^2 \delta\left(\epsilon_n -
E^{\eta}_{\nu,k}(\vec{\Omega})\right)
\end{equation}
where $|k\nu\rangle= \sum_{\sigma} {\cal
C}_{k,\nu,\sigma}(\vec{\Omega})|\sigma\rangle|k\rangle$. It is apparent that
$\Gamma_{n\sigma}$ depends on $\vec{\Omega}$ both through the density of states
and throug the mixing coefficients   $  {\cal C}_{k,\nu,\sigma}$. The
tunability of $\Gamma$ can bring and additional knob to study the SET in the
Kondo regime \cite{Ralph-Science}.


In summary, we have provided a simple conceptual framework to
understand the different effects that occur in a non-magnetic single electron
island coupled to ferromagnetic electrodes with magnetic anisotropy.
The sensitivity of a single-electron transistor device to the
chemical potential of the electrodes results in  new physical effects  when
these are ferromagnetic.  Both the single particle lifetimes $\Gamma$
and the charge vs gate curve depend on the orientation of the magnetic moment
 $\vec{\Omega}$ with
respect to the easy axis. This effect can be used to probe the chemical
potential of ferromagnetic electrodes and could have 
practical applications. Further theoretical 
work is necessary to provide realistic description of the electrodes as
well as to consider the case of ferromagnetic island, either
metallic \cite{Canali}
 or semiconducting \cite{JFR04}.

\begin{acknowledgement}
We acknowledge fruitful discussions with J. J. Palacios, D. Jacob
and C. Tejedor.  JFR acknowledges funding from Generalitat
Valenciana (GV05-152), Spanish  MEC 
(Grants FIS200402356) and Programa Ram\'on y Cajal.  
RA and LB acknowledge funding from the Spanish MEC (grant
MAT2005-07369-C03-03).

\end{acknowledgement}

\end{document}